\documentclass[aps,prb,twocolumn,floatfix,groupedaddress,showkeys]{revtex4}
\usepackage{graphicx}   
\usepackage{bm}

\begin{document}

\title{Projective Dynamics in Realistic Models of Nanomagnets}

\author{S.H.\ Thompson}\email[]{sam@csit.fsu.edu}
\affiliation{Department of Physics, Center for Materials Research and 
Technology, and School for Computational Science and Information Technology, 
Florida State University, Tallahassee, FL 32306-4350, USA}

\author{G.~Brown}\email[]{browngrg@csit.fsu.edu}
\altaffiliation[Also at]{ CSIT, Florida State University, Tallahassee, FL 32306-4120, USA}
\affiliation{Center for Computational Sciences, 
Oak Ridge National Laboratory, Oak Ridge, TN 37831-6164, USA}

\author{P.A.\ Rikvold}\email[]{rikvold@csit.fsu.edu}
\affiliation{Department of Physics, Center for Materials Research and 
Technology, and School for Computational Science and Information Technology, 
Florida State University, Tallahassee, FL 32306-4350, USA}

\date{\today}

\begin{abstract}
The free-energy extrema governing the magnetization-reversal process for a model 
of an iron nanopillar are investigated using the projective dynamics method.
Since the time evolution of the model is computationally intensive, one
could question whether sufficient statistics can be generated, with current
resources, to evaluate the position of the metastable configuration.  
Justification for the fine-grained discretization of the model that we use here is given, 
and it is shown that tractable results can be obtained for this system on realistic time scales.
\end{abstract}


\maketitle

\section{Introduction}
\label{sec:I}
Magnetic nanopillars can be fabricated \cite{1}, which offer 
interesting opportunities for direct comparison with numerical models.  
In the application discussed here, the pillars are grown perpendicular to a surface.
For an applied field parallel to the long axis of the pillar, there exists
a stable free-energy minimum when the field is parallel to the average
magnetization of the pillar, and a metastable minimum for the antiparallel
orientation.  
Magnetization reversal involves a transition from the metastable minimum to the 
stable minimum across a saddle point in the free energy.  
It is the free-energy difference between the saddle point and the metastable 
minimum that determines the switching time of the pillar.  
While determining the magnetization of the metastable minimum is comparatively 
easy, the magnetization of the  saddle point is more difficult to determine.  
Projective Dynamics (PD)  is a technique that has proven capable of 
finding the saddle point in other magnetization-reversal systems \cite{3,2}.

\section{Model and Numerical Results}
\label{sec:M}
In order for the proper dynamics to emerge, the magnetization of the simulated 
pillar must be discretized on a sufficiently fine scale.  
Pillar models that have previously been studied by projective dynamics consisted 
only of a one-dimensional chain of spins \cite{2}. 
This model lacks richer dynamics, such as magnetization curling modes, that are seen in
experiments and in more realistic models.  
In addition, the dependence of the switching field on the angle of misalignment
between the pillar axis and the applied field observed in the one-dimensional model \cite{5}
does not correspond to that seen in the experimental pillars \cite{4,8,9}.

To achieve more realistic dynamics, a model pillar with physical dimensions 
10 nm $\times$ 10 nm $\times$ 150 nm was discretized onto a 6 $\times$ 6 $\times$ 
90 cubic lattice.  
The model parameters were chosen to be consistent with bulk iron: exchange length of
2.6 nm and magnetization density of 1700 emu/cm$^3$.
The temperature was fixed at 20 K.

The dynamics of the system are governed by the Landau-Lifshitz-Gilbert (LLG)
equation \cite{7}, 
\begin{equation}
\frac{\rm{d}\vec{M}(\vec{r_i})}{\rm{dt}} = \frac{\gamma_0}{1+\alpha}\left(\vec{M}(\vec{r_i})\times 
\left[\vec{H}(\vec{r_i})-\frac{\alpha}{M_s}\vec{M}(\vec{r_i})\times \vec{H}(\vec{r_i})\right]\right),
\label{eq:M}
\end{equation}
which describes the time evolution of the magnetization in the presence of a local field.
Here $\gamma_0$ is the electron gyromagnetic ratio with a value of 
$1.67\times10^{7}$ Hz/Oe, and $\alpha$ is a phenomenological damping parameter, 
chosen as 0.1 to give underdamped behavior.  
$\vec{M}(\vec{r_i})$ is the magnetization density at position $\vec{r_i}$, with a constant
magnitude $M_s$,  and $\vec{H}(\vec{r_i})$ is the total local field at $\vec{r_i}$. 
The latter contains contributions from the dipole-dipole interactions, exchange
interactions, and the applied field.

A reversal simulation begins with the magnetization oriented along the long axis of 
the pillar in the $+z$ direction.
It is allowed to relax in the presence of a field $+B_0\hat{z}$ before the field is quickly
switched to $-B_0\hat{z}$, where $B_0 = 1125$~Oe has been chosen smaller than the 
coercive field.
\begin{figure}[t]
\centering
\vspace{0.5truecm}
\includegraphics[width=.47\textwidth]{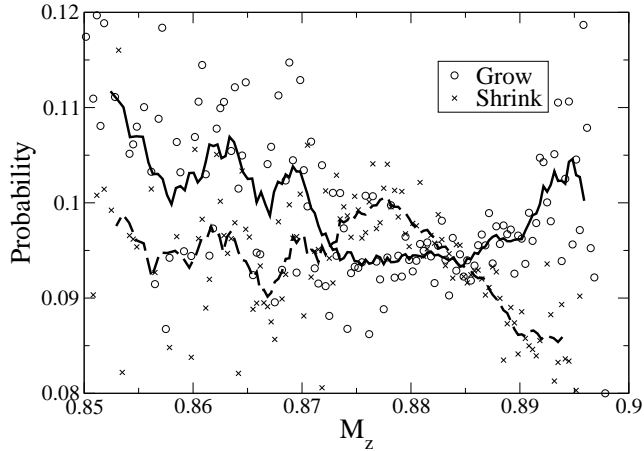}
\caption[]{
Growing and shrinking probabilities versus the magnetization along the $z$-axis for a
6$\times$6$\times$90 pillar.  
Grow/shrink data were collected into 6000 bins, which span the entire magnetization axis, [-1,1].
The simulation was run at 20 K, 1125 Oe, and the applied field was directed along the
long axis of the pillar, antiparallel to the initial magnetization.  
The continuous curves represent nine-point running averages
}
\label{fig:fig1}
\end{figure}

Projective dynamics involves analyzing the probability of growing toward the stable
equilibrium or shrinking away from it, as a function of a single variable that measures
the progress of the reversal process.
To collect growing and shrinking statistics here, the $z$-component
of the total magnetization of the pillar, $M_z$, was recorded at each integration step.  
In addition, the change in this value from the previous integration step, $\Delta M_z$,
was also recorded.  
The magnetization axis was discretized into bins into which the collected data were sorted.  
The bin size was determined such that the discretized $\Delta M_z$ would only involve changes
between adjacent bins.
The growing and shrinking probabilities were calculated by counting the number of times
that $\Delta M_z$ would cause the magnetization to change bins.  
Define $G$ as the number of times the magnetization hopped to the next bin in the direction
toward equilibrium, and $S$ as the number of hops in the direction back toward the metastable
state.
If $N$ is the total number of visits to the starting bin, then the growing and shrinking probabilities
for each bin, $P_G$ and $P_S$, can be written respectively as

\begin{equation}
P_G = \frac{G}{N},\qquad 
P_S = \frac{S}{N}.
\label{eq:P}
\end{equation}

The crossings of the growing and shrinking probabilities give information about the
locations of the metastable free-energy minimum and the saddle point.  
Where the two probabilities are equal, the system has no preferred direction of
magnetization evolution, indicating that the free-energy landscape is flat in this region.
This is the location of an extremum.
The right-hand intersection in Fig.~\ref{fig:fig1} corresponds to the metastable free-energy minimum, 
where the system spends most of its time.   
At the saddle point, the growing and shrinking probabilities are also equal, as indicated
by the left-hand crossing.
The region between these two crossings shows a higher probability for shrinkage than
growth, corresponding a preference for moving toward the metastable minimum.

Previous simulations performed on a one-dimensional chain of seventeen spins 
exhibited Stoner-Wohlfarth behavior for the angular dependence of the switching field \cite{5}. 
This is not representative of experimental data \cite{4,8,9}.  
However, a full three-dimensional model qualitatively produced the proper angular 
dependence \cite{5}.
This is not surprising, given that the full three-dimensional model allows for magnetization curling
modes which are simply not allowed in the one-dimensional model. 
PD has previously been applied to the one-dimensional spin chain \cite{2}, but several thousand switches 
had to be performed in order to accumulate sufficient statistics to reveal the free-energy 
extrema \cite{2}.  
With typical switches taking around 1500~CPU-hours for the full three-dimensional model,
it was not clear that sufficient statistics could be generated. 

Figure \ref{fig:fig1} shows the results collected from only 20 switches for the full model.  
The estimated $P_G$ is shown as circles, and $P_S$ as crosses.
Scatter in these data is primarily due to the counting statistics and the small number
of samples in each bin.
Nine-point running averages are used to reduce the scatter and improve estimates of
the crossings of $P_G$ and $P_S$.
The running averages are represented by curves and are sufficient to
determine the locations of the extrema with an uncertainty of about $1\times10^{-3}$ in 
the magnetization.  
For the one-dimensional model the estimated curves representing $P_G$ and $P_S$ are 
almost parallel near the saddle point, making it difficult to locate crossings of the growing 
and shrinking probabilities (see Fig.~1 of Ref.~\cite{2}).
Since they cross at a significant angle for the model studied here (see Fig. \ref{fig:fig1}), it is 
possible to find an estimate for the magnetization of the free-energy saddle point from 
far fewer switches than were needed for the one-dimensional model.

\section{Summary and Conclusions}
\label{sec:S}
The projective dynamics method was used to produce growing and shrinking probabilities 
during the magnetization reversal of a realistic model nanopillar.  
These probabilities can be used to locate the magnetization of the free-energy extrema. 
In particular, the magnetization of the saddle point is revealed. 
Compared to the one-dimensional model, far fewer switches are needed to find the location
of the extrema in the three-dimensional model.
In the future, a study of the dependence of the saddle
point on temperature and misalignment between the pillar and the applied field are planned.

\section*{Acknowledgments}
This work was supported by U.S.~National Science Foundation Grant No.~DMR-0120310, 
and by Florida State University through the School of Computational Science and Information
Technology and the Center for Materials Science and Technology.
This work was also supported by the DOE Office of Science through ASCR-MICS and the
Computational Materials Science Network of BES-DMSE, as well as the Laboratory directed
Research and Development program of ORNL under contract DE-AC05-00OOR22725 with
UT-Battelle LLC.

\end{document}